# Tunable spin-phonon polarons in a chiral molecular qubit framework

Aimei Zhou[1,2†], Ruihao Bi[1,2†], Zhenghan Zhang[3], Luming Yang[4], Xudong Tian[3], Denan Li[5], Yingchao Wang[1,2], Mingshu Tan[6], Weibin Ni[2,5], Haozhou Sun[1,2], Jinkun Guo[7], Xiaohe Miao[8], Xinxing Zhao[9], Zhifu Shi[9], Wei Tong[10], Zhitao Zhang[10], Jiandong Feng[11], Jin-Hu Dou[7], Feng Jin[6], Shi Liu[1,2,5], Mircea Dincă[4], Tijana Rajh[12,13], Jian Li[3], Wenjie Dou[1,2,5*], Lei Sun[1,2,5*]

[1] Department of Chemistry, School of Science and Research Center for Industries of the Future, Westlake University, Hangzhou, Zhejiang Province 310030, China

[2] Institute of Natural Sciences, Westlake Institute for Advanced Study, Hangzhou, Zhejiang Province 310024, China

[3] State Key Laboratory of Coordination Chemistry, School of Chemistry and Chemical Engineering, Nanjing University, Nanjing, Jiangsu Province 210023, China

[4] Department of Chemistry, Massachusetts Institute of Technology, Cambridge, Massachusetts 02139, United States

[5] Department of Physics, School of Science and Research Center for Industries of the Future, Westlake University, Hangzhou, Zhejiang Province 310030, China.

[6] Beijing National Laboratory for Condensed Matter Physics, Institute of Physics, Chinese Academy of Sciences, Beijing 100190, China

[7] National Key Laboratory of Advanced Micro and Nano Manufacture Technology, School of Materials Science and Engineering, Peking University, Beijing 100871, China

[8] Zhejiang Key Laboratory of Precise Synthesis of Functional Molecules, Instrumentation and Service Center for Molecular Sciences and Research Center for Industries of the Future, Westlake University, 600 Dunyu Road, Hangzhou 310030, Zhejiang Province, P. R. China

[9] CIQTEK Co., Ltd., Hefei, Anhui Province 230021, China

[10] Anhui Key Laboratory of Low-Energy Quantum Materials and Devices, High Magnetic Field Laboratory, HFIPS, Chinese Academy of Sciences, Hefei, Anhui 230031, China

[11] Laboratory of Experimental Physical Biology, Department of Chemistry, Zhejiang University, 310058 Hangzhou, China

[12] Center for Nanoscale Materials, Argonne National Laboratory, Argonne, Illinois 60439, United States

[13] The School of Molecular Sciences, Arizona State University, Tempe, Arizona 85281, United States

[†]These authors contributed equally to this work.

[*]Email: sunlei@westlake.edu.cn; douwenjie@westlake.edu.cn

## Abstract

Chiral structures that produce asymmetric spin-phonon coupling can theoretically generate spin-phonon polarons—quasiparticles exhibiting non-degenerate spin states with phonon displacements. These quasiparticles are speculated to be the origin of chirality-induced spin selectivity and presumably can display exotic dynamic behaviors. However, direct experimental evidence of spin-phonon polarons has been lacking. Using a chiral molecular qubit framework embedding stable semiquinone-like radicals, we report spin dynamic signatures that indicate the formation of spin-phonon polarons for the first time. Our non-adiabatic model reveals that these quasiparticles introduce an active spin relaxation channel when polaron reorganization energy approaches Zeeman splitting. This new channel manifests itself as anomalous, temperature-independent spin relaxation, which can be suppressed by high magnetic fields or pore-filling solvents (e.g. $CH_2Cl_2$, $CS_2$). Such field- and guest-tunable relaxation is unattainable in conventional spin systems. Harnessing this mechanism could boost repetition rates in spin-based quantum information technologies without compromising coherence or quantum sensing performance.

Chiral structures lack inversion and reflection symmetries. They exhibit reciprocity with electron spins: passing electrons through chiral biomolecules[1], polymers[2], and inorganic solids[3] could generate salient spin polarization, an effect called chirality-induced spin selectivity (CISS)[4], and spin-polarized current could induce enantioselective electrochemical reactions[2]. These intriguing phenomena hold promise for new designs and methodologies towards high-performance spintronic devices[5], quantum information science[6], and asymmetric synthesis[7]. Nonetheless, the underlying mechanism of the interplay between structural chirality and electron spin degree of freedom is still an open question[8]. Recent theoretical studies indicate that chiral structures could generate asymmetric spin-phonon coupling, which infers the formation of a polaron-like bound state with energy splitting and phonon displacement between opposite spin states[9,10]. These quasiparticles, which we name spin-phonon polaron, are speculated to be the origin of many exotic phenomena including the CISS effect[9,10], yet experimental evidence of their presence is elusive. As spin-phonon coupling also drives spin relaxation and implicitly decoherence[11–13], the spin-phonon polaron should manifest themselves with peculiar spin dynamics, which remains unexplored both theoretically and experimentally.

To this end, molecular qubit frameworks (MQFs) provide a tunable platform to investigate spin-phonon polarons and their dynamics. MQFs integrate molecular electron spin qubits within open framework materials, such as metal−organic frameworks[14,15] and covalent organic frameworks[16]. Their highly ordered crystal structures engender well-defined phonon dispersion relations, their bottom-up synthesis allows rational design of spin, phonon, and chirality characteristics, and their microporosity and high surface area further allow fine tuning of phonons by harnessing host-guest interactions[17–19]. In addition, when stable organic radicals are used as qubits, MQFs could maintain quantum coherence at room temperature, enabling spin dynamic characterization across a wide range of temperature (3 K – 300 K)[20–22]. This empowers investigation of spin-phonon coupling involving acoustic and optical phonons that stem from long-range structural features and local chemical bonds, respectively[11,12]. Hence, MQFs with chiral structures and radical qubits could enable systematic examination of spin relaxation through rational structural design and phononic modulation.

Herein, we report spin dynamic signatures that clearly indicate the formation of spin-phonon polarons in a chiral and microporous MQF, $Zn_3(HOTP)$ (HOTP = 2,3,6,7,10,11-hexaoxytriphenylene). Spontaneous oxidation of HOTP ligands generates semiquinone-like radical qubits with room-temperature electron spin coherence. They display a fast and nearly temperature-independent spin relaxation process at 0.34 T, which is suppressed by increasing the magnetic field or filling the pores with some organic solvents including dichloromethane (DCM), tetrahydrofuran (THF), and carbon disulfide ($CS_2$). Considering that the chiral structure could induce asymmetric longitudinal spin-phonon coupling, we build up a non-adiabatic microscopic model that allows for the formation of spin-phonon polaron and solve it with polaron transformation as well as Fermi's golden rule similar to the manner of Marcus rate theory[23–26]. This model shows that the anomalous and tunable spin relaxation of $Zn_3$(HOTP) arises from an additional relaxation channel introduced by spin-phonon polarons, which is effective when the polaron reorganization energy approaches Zeeman splitting of different spin states.

## Chiral structure with incommensurate modulation

$Zn_3$(HOTP) was synthesized via a solvothermal reaction between Zn(acetate)$_2\cdot 2H_2O$ and 2,3,6,7,10,11-hexahydroxytriphenylene (HHTP), which yielded hexagonal rod-like crystals with longitudinal lengths of 30 – 70 μm and cross-sectional lengths of 1 – 2 μm (Supplementary Fig. 1). The product is crystalline, air-stable, and microporous with a Brunauer−Emmett−Teller surface area of 362 $m^2\cdot g^{-1}$ (Supplementary Figs. 2 and 3) that is comparable with a previously reported value[27]. Its structure remains intact in various organic solvents, e.g. DCM, THF, and *N*, *N*-dimethylformamide (DMF), at both 295 K and 100 K (Supplementary Figs. 3 – 5, Supplementary Table 1). Individual single crystal of $Zn_3$(HOTP) displays second harmonic generation (SHG) activity at 295 K and 100 K (Fig. 1d, Supplementary Fig. 6), indicating spontaneous inversion symmetry breaking in this material[28].

Single-crystal structure of $Zn_3$(HOTP) was characterized by continuous rotation electron diffraction (cRED) at 96 K and further refined by extended X-ray absorption fine structure (EXAFS) spectroscopy (Fig. 1a, Supplementary Figs. 7 and 8, Supplementary Tables 2 and 3). Based on the SHG activity of this framework, we solved its structure with non-centrosymmetric space groups and found that it crystalizes in $P6_3$, a Sohncke space group, with a chiral structure (Fig. 1f)[29]. $Zn_3$(HOTP) exhibits a honeycomb-like non-van-der-Waals three-dimensional (3D) framework with cylindrical pores whose diameters are approximately 1.6 nm (Fig. 1e). HOTP ligands form eclipsed-stacked columns along the crystallographic *c* axis, which aligns with the long axis of single crystal as revealed by high-resolution transmission electron microscopy (HR-TEM), with an intermolecular distance of 3.305 Å (Fig. 1c and 1e). These columns are connected via zigzag Zn−O ladders constructed by [$Zn_2O_2$] pseudo-squares where $Zn^{2+}$ ions are not co-planar with HOTP ligands (Fig. 1e, Supplementary Fig. 9). Each $Zn^{2+}$ displays a trigonal bipyramidal coordination geometry with an axial water pointing into the pores (Fig. 1e). Notably, $Zn_3$(HOTP) is structurally and compositionally distinct from $Zn_3(HOTP)_2$, an alternative phase recently reported by Dincă et al[30]. The latter, which is produced by a different synthetic procedure, consists of 4- and 5-coordinated $Zn^{2+}$ ions without coordinating water.

Electron diffraction of $Zn_3$(HOTP) along the crystallographic *c* axis displays satellite reflections at 96 K (Fig. 1a). Their periodicity is approximately 3.57 times of the corresponding unit cell parameter, yet it is irrational relative to the latter, characteristic of incommensurate modulation. Such structural aperiodicity has also been observed in several other HOTP-based MOFs including $Zn_3(HOTP)_2$[30], $La_3(HOTP)_2$, and $Nd_3(HOTP)_2$[31]. It reduces $6_3$ screw axes to 3-fold rotation axes while retains structural chirality (Supplementary Fig. 10). $Zn_3$(HOTP) transforms into a periodic structure at 295 K (Fig. 1b), yet it still exhibits SHG activity (Fig. 1d), showing that the chirality is intrinsic to this structure.

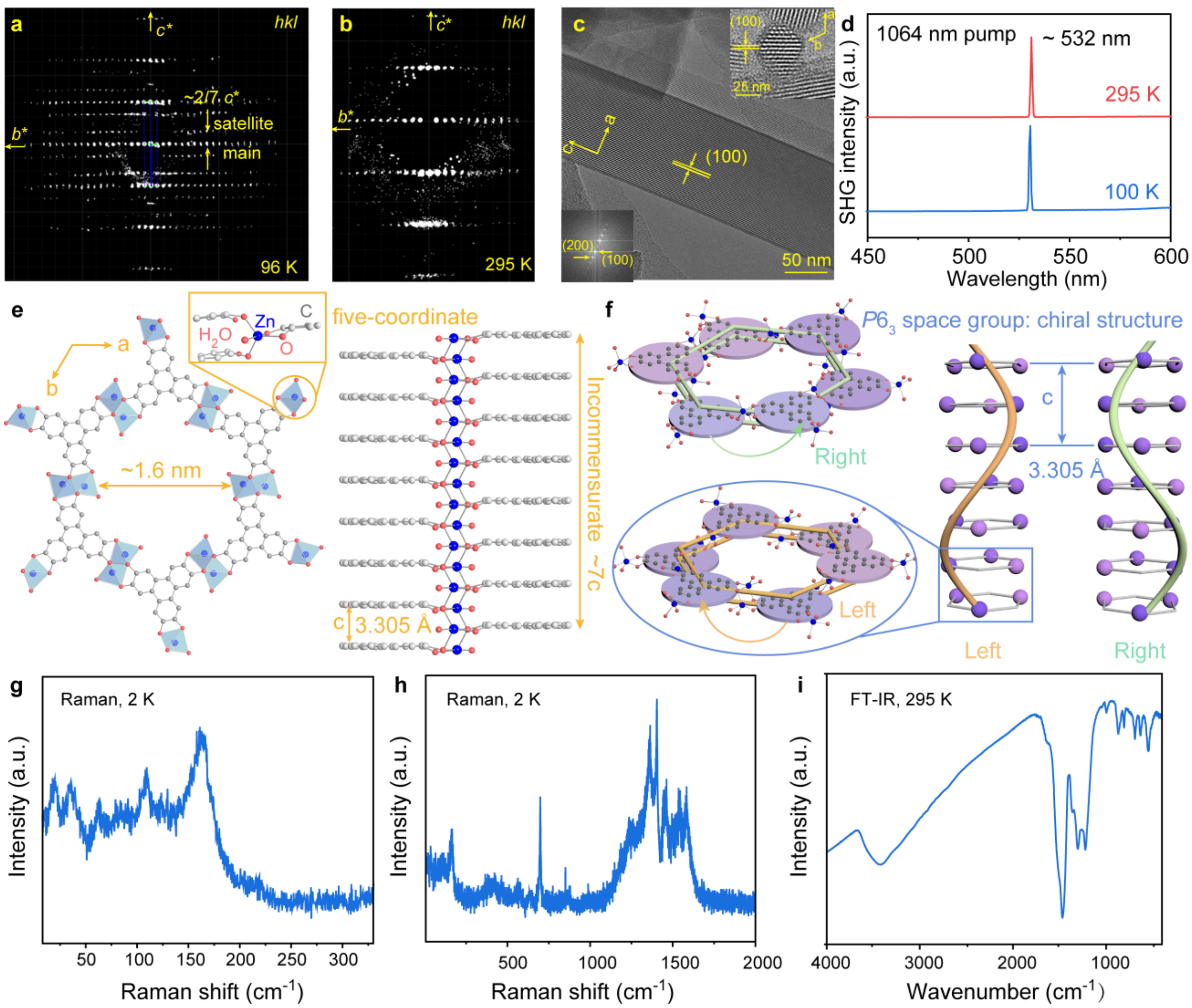

**Fig. 1. Structure and optical phonons of $Zn_3$(HOTP). a,b,** cRED images in reciprocal space acquired at (**a**) 96 K and (**b**) 295 K. Red, green, and blue lines represent $a^*$, $b^*$, and $c^*$, respectively. Main and satellite reflections are labeled. **c,** HR-TEM image viewed perpendicular to the crystallographic $c$ axis. Bottom-left inset: Fourier transform of the HR-TEM image. Top-right inset: HR-TEM image viewed along the crystallographic $c$ axis. **d,** SHG signals under 1064 nm pumping acquired at 100 K and 295 K. **e,** Portions of the structure viewed along and perpendicular to the crystallographic $c$ axis. Inset: coordination environment of $Zn^{2+}$. Blue, grey, and red spheres represent Zn, C, and O, respectively. H atoms are omitted for clarity. **f,** Left: Double-layer crystal packing showing structural chirality. Each circular plate represents a HOTP ligand and nearby secondary building units. They are marked with different colors to articulate the $6_3$ screw-axis symmetry. Yellow and green arrows indicate clockwise and counterclockwise rotations corresponding to left and right handedness, respectively. Right: Chiral structures with left and right handedness in $P6_3$ space group viewed perpendicular to the crystallographic $c$ axis. **g,h**, Raman spectra collected at 2 K under 0.34 T. **i**, FT-IR spectrum collected at 295 K under 0 T.

We employed Raman and Fourier-transform infrared (FT-IR) spectroscopy for $Zn_3$(HOTP) to investigate phonons near the Gamma point of $1^{st}$ Brillouin zone. Vibrational peaks at 3420 $cm^{-1}$ and 699 $cm^{-1}$ are assigned to stretch modes of O−H and Zn−O bonds, respectively, and those at 1200 $cm^{-1}$ − 1600 $cm^{-1}$ are attributed to C−C and C−O stretches within HOTP (Fig. 1g–i). Raman spectrum collected at 2 K and 0.34 T shows multiple peaks below 200 $cm^{-1}$ with the lowest-frequency peak at 15 $cm^{-1}$ (Fig. 1g). These low-frequency Raman features are reproduced by calculations, indicating that they are inherent for the framework (Supplementary Fig. 11b). They pose the upper limit to Debye temperature ($T_D$ < 22 K), indicating relatively low structural rigidity. A strong peak at 163 $cm^{-1}$ red-shifts to 151 $cm^{-1}$ for $Zn_3$(HOTP) synthesized in $D_2O$ (Supplementary Fig. 11a), showing its stemming from hydrogen bonds formed between adsorbed $H_2O$ and HOTP or between adsorbed and coordinating $H_2O$. As the electron spin of HOTP radical mainly resides on its oxygen and carbon atoms (Supplementary Fig. 12)[22], vibrations of hydrogen bonds between HOTP and $H_2O$, Zn−O bonds, C−C bonds, and C−O bonds may contribute to local-mode spin relaxation in $Zn_3$(HOTP).

## Conventional polarons and spin coherence

We probed the spin characteristics of $Zn_3$(HOTP) with continuous-wave electron paramagnetic resonance (CW-EPR) spectroscopy. The X-band (9.6 GHz, 0.34 T) CW-EPR spectrum acquired at 295 K reveals two types of $S$ = 1/2 electron spins exhibiting Lorentzian line shape: one at $g$ = 2.00341 with a linewidth of 0.40 mT and the other at $g$ = 2.00237 with a linewidth of 2.37 mT (Fig. 2a). Both $g$-factors are close to the free-electron value ($g_e$ = 2.00232) and consistent with radical characteristics. Power-saturation experiments revealed $T_1$ = 330 ns and $T_2^*$ = 33 ns for the first type and $T_1$ = 27 ns and $T_2^*$ = 5 ns for the second type at room temperature ($T_1$ and $T_2^*$ represent spin relaxation time and spin dephasing time, respectively) (Supplementary Fig. 13). As the temperature rises from 130 K to 295 K, the double integration, which is proportional to magnetic susceptibility ($\chi$) of the first type, drops significantly (Fig. 2b and c). This is consistent with Curie-like paramagnetism[32,33]. In contrast, the second type displays nearly temperature-independent paramagnetism (TIP). Q-band (34 GHz, 1.22 T) and millimeter-wave (214 GHz, 7.64 T) CW-EPR spectra acquired at 295 K exhibit features with slight $g$-anisotropy and narrow linewidths (Supplementary Fig. 14). Signals originating from the second type of spins were not discernable at high magnetic fields.

The drastically different linewidths and paramagnetic properties of the two types of electron spins suggest their distinct physical origins. These behaviors are comparable with previous observations of doped polyaniline[32] and fully reduced $Cu_3(C_6O_6)_2$ ($H_4C_6O_6$ = tetrahydroxy-1,4-benzoquinone)[33], where the Curie-like and TIP-type spins were assigned to localized defects and delocalized electrons, respectively. The latter should contribute to electrical conduction whereas the former should not. Single-crystal electrical characterization of $Zn_3$(HOTP) revealed an electrical conductivity of 0.46 $S{\cdot}cm^{-1}$ along the crystallographic $c$ axis, verifying the presence of delocalized electrons and their efficient transport through π-stacked HOTP columns (Supplementary Figs. 15 and 16, Supplementary Table 4). Therefore, although the physical origin of TIP remains unclear and under debate, we tentatively assign it to delocalized electrons, whose formation is facilitated by the π−π stacking along the crystallographic $c$ axis, and attribute the Curie-like spins to localized defects, i.e. semiquinone-like radicals formed from spontaneous oxidation of HOTP in the air (Supplementary Fig. 17). These species should couple with phonons to form large (Fröhlich) polarons and small (Holstein) polarons, respectively[34]. Density functional theory (DFT) calculation indicates that the oxidized HOTP defects, albeit possessing different oxidation states from the majority unoxidized ligands, do not induce significant structural distortion (Supplementary Fig. 12). Quantitative EPR analysis revealed that approximately 0.99% of HOTP ligands form small polarons (Supplementary Table 5).

Our previous studies have shown that localized HOTP radicals in MQFs behave as electron spin qubits with room-temperature quantum coherence[15,22]. This prompted us to investigate the spin dynamics of $Zn_3$(HOTP) and to explore its potential as an MQF with X-band pulse EPR spectroscopy. At 100 K, 180 K, and room temperature, the echo-detected field sweep spectra display single peaks with Gaussian linewidths below 0.52 mT (Supplementary Fig. 18). Hence, only the small polarons can be probed, whereas the large polarons lose coherence so fast that they are not measurable. Nutation experiments conducted at 295 K revealed Rabi oscillations, demonstrating coherent addressability (Supplementary Fig. 19). The phase memory time ($T_m$), which reflects all dephasing processes and represents the spin decoherence time ($T_2$) for ensemble systems, is approximately 700 ns at 13 K. It increases to 1.01 μs at 63 K and gradually decreases to 376 ns at 295 K (Fig. 2d). The latter value is an order or magnitude longer than the room-temperature $T_2^*$ of small polarons because it was acquired by the Hahn echo decay sequence that dynamically decouples the electron spins from environmental magnetic noise[11]. Overall, the small polarons of $Zn_3$(HOTP) behave as room-temperature electron spin qubits, qualifying this material as an MQF.

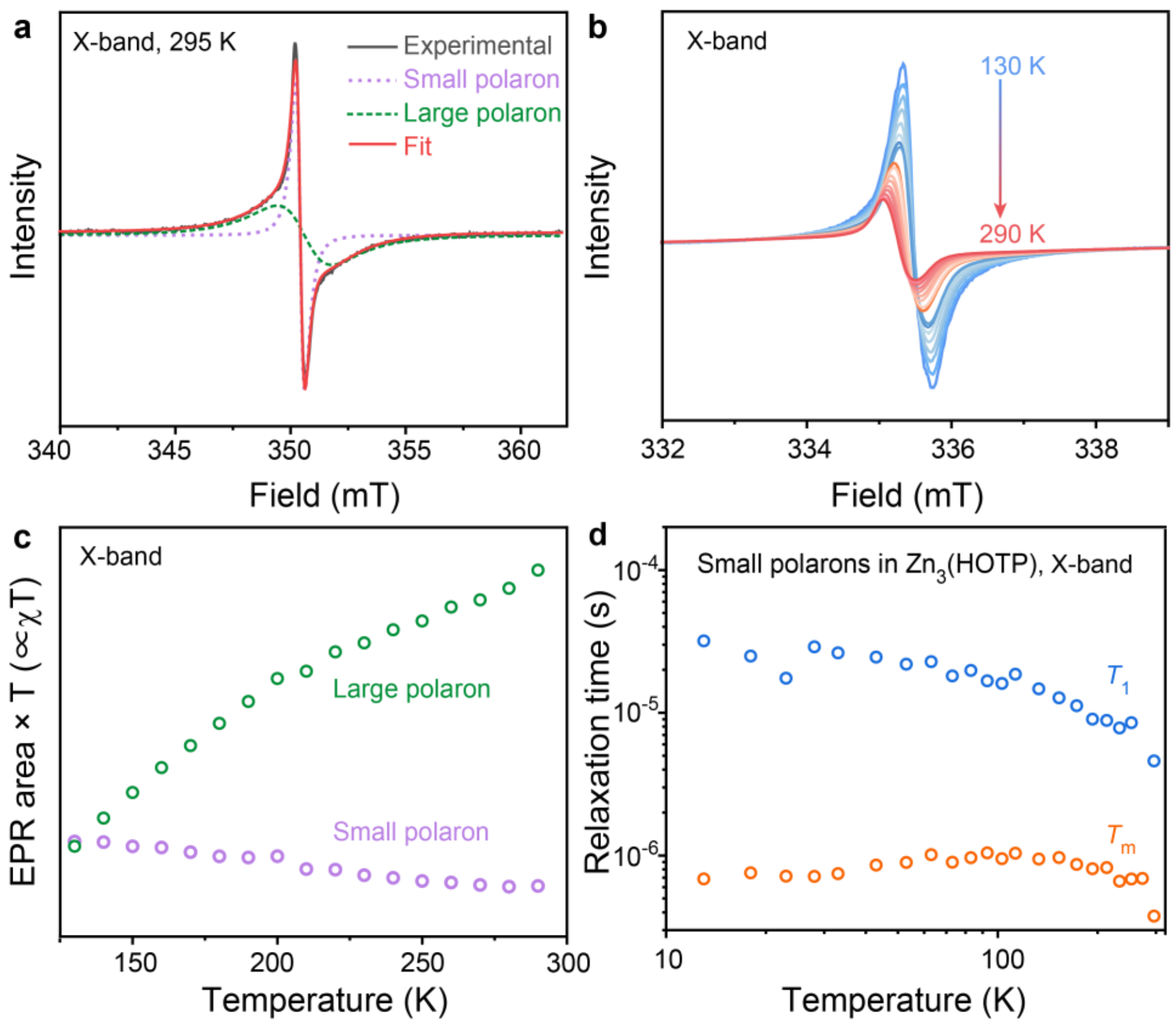

**Fig. 2. CW-EPR spectra of $Zn_3$(HOTP). a,** CW-EPR spectrum collected at X-band and 295 K and associated fitting results. **b,** CW-EPR spectra collected at X-band from 130 K to 290 K. **c,** EPR area × T (proportional to $\chi$T) vs temperature (T) data for large and small polarons. **d,** $T_1$ and $T_m$ of small polarons collected at X-band with inversion recovery and Hahn echo decay sequences, respectively, from 13 K to 295 K. Error bars are omitted for clarity as they are smaller than the data point symbols (Supplementary Table 6).

**Anomalous and tunable spin relaxation of small polarons**

$Zn_3$(HOTP) exhibits exceptionally short $T_1$ at X-band with weak temperature dependence from 13 K to 295 K. $T_1$ is 31.9 μs at 13 K, which is shorter than the value of another HOTP-based MQF, $[(CH_3)_2NH_2]_2$Ti(HOTP), at the same temperature by three orders of magnitude[22]. It gradually decreases as the temperature rises, reaching nearly 4 μs at 295 K (Fig. 2d). This spin relaxation behavior differs from all other stable organic radical qubits[11,21]. To articulate spin relaxation mechanisms of $Zn_3$(HOTP), we further acquired its $T_1$ at various temperatures with Q-band (1.22 T, 34 GHz) and W-band (3.34 T, 94 GHz) pulse EPR spectroscopy (Fig. 3a and b). Raising the magnetic field significantly suppresses spin relaxation below 100 K and induces a significant temperature dependence of $T_1$ (Fig. 3c). Specifically, $T_1$ became 3.39 ms and 1.59 ms at 13 K under 1.22 T and 3.34 T, respectively, both of which are nearly two orders of magnitude longer than the value observed under 0.34 T. The spin relaxation rates ($1/T_1$) collected at Q-band and W-band exhibit comparable temperature dependencies above 18 K, yet they diverge below this temperature with $T_1$ being slightly shorter under higher magnetic field. This comparison indicates contributions of a field-dependent direct process and field-independent two-phonon (Raman and local-mode) processes to the spin relaxation in $Zn_3$(HOTP) at Q-band and W-band[12,22].

Based on these results, we tentatively analyzed spin relaxation mechanisms of $Zn_3$(HOTP) with the following equation:

$$\frac{1}{T_1} = C + A_{Dir}T + \sum_i A_{Loc,i}\frac{e^{h\nu_i/k_BT}}{\left(e^{h\nu_i/k_BT}-1\right)^2} \quad (1)$$

where the three terms describe a temperature-independent spin relaxation process, the direct process, and local-mode processes. C is a constant; ν, $k_B$, and T represent linear vibrational frequency, Boltzmann constant, and temperature, respectively. $A_{Dir}$ and $A_{Loc}$ are pre-factors of direct and local-mode processes, respectively. Fitting the temperature dependence of $1/T_1$ reveals that the spin relaxation at X-band involves a temperature-independent process and a local-mode process driven by the vibration of hydrogen bonds ($\nu/c = 163$ $cm^{-1}$; c represents the speed of light; Supplementary Fig. 20). The former is dominant below 170 K with an exceptionally high rate ($C = 4.5 \times 10^4$ $s^{-1}$) that indicates a fast spin relaxation, and the latter causes a slight decrease of $T_1$ towards room temperature. At Q-band and W-band, the direct process plays a major role below 50 K, and local-mode processes driven by vibrations of hydrogen bonds and Zn−O bonds ($\nu/c = 699$ $cm^{-1}$) dominate at higher temperatures (Fig. 3e and f; Supplementary Table 17). Notably, the temperature-independent spin relaxation process is quenched at these magnetic fields, indicating that this process is field-tunable and sensitive to the Zeeman splitting of small polarons (Fig. 3b).

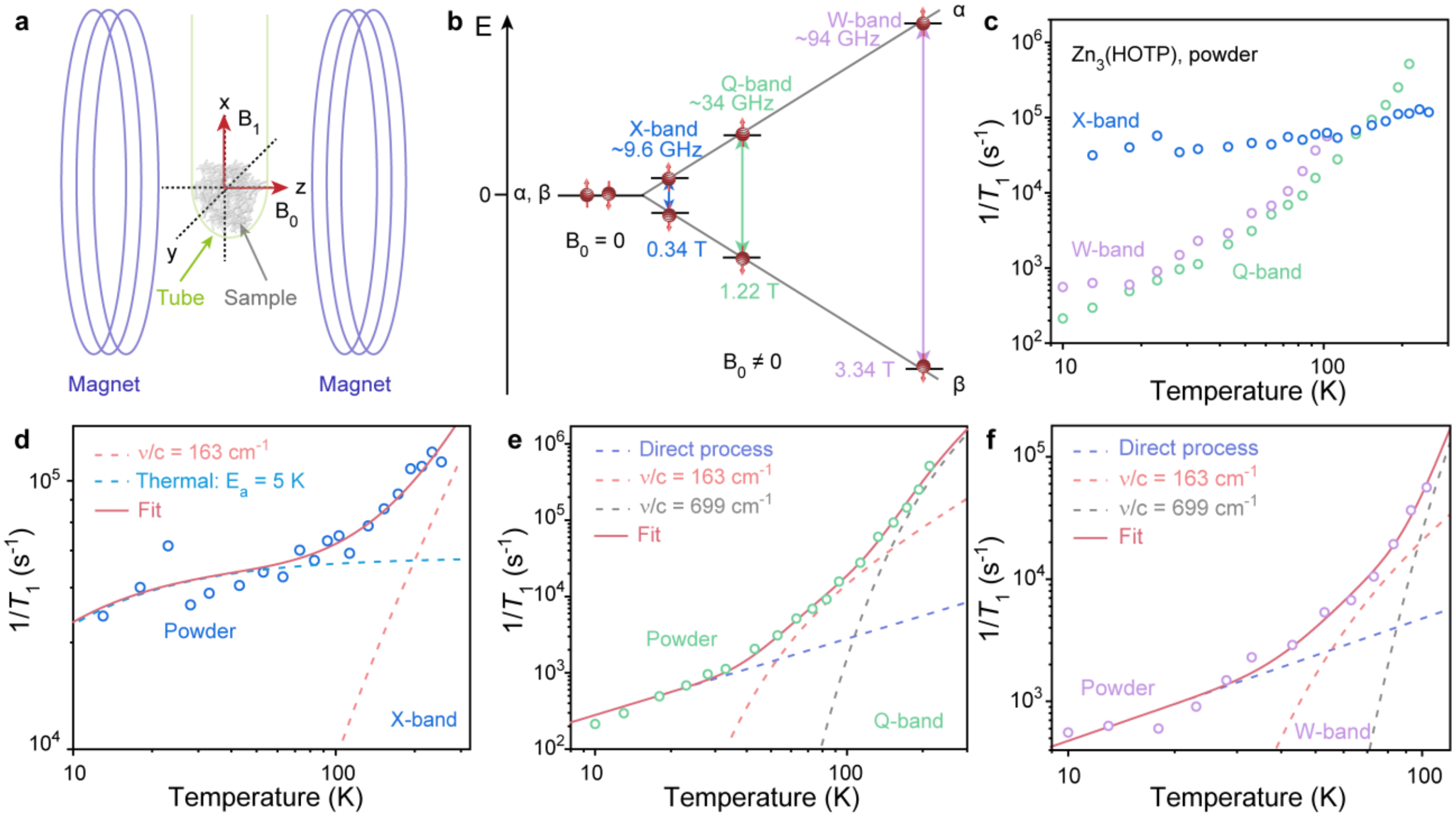

**Fig. 3. Magnetic field modulation of spin relaxation. a,** Schematic diagram of a sample of $Zn_3$(HOTP) crystallites in an EPR spectrometer. $B_0$ represents a static magnetic field and $B_1$ the microwave magnetic field. **b,** Zeeman splitting of small polarons in $Zn_3$(HOTP). Magnetic fields and corresponding microwave frequencies used in experiments are indicated. The red sphere with an arrow represents an electron spin. E represents energy. α and β represent different spin states. **c,** Temperature dependence of $1/T_1$ for $Zn_3$(HOTP) powders acquired at X-band, Q-band, and W-band. Error bars are omitted for clarity as they are smaller than the data point symbols (Supplementary Tables 8 and 9). **d−f**, Spin relaxation mechanisms of $Zn_3$(HOTP) powders at (**d**) X-band, (**e**) Q-band, (**f**) W-band. Dash lines labeled as "Thermal" and "υ/c" represent thermally activated and local-mode processes, respectively.

We further probed the influence of phonons on the spin relaxation in $Zn_3$(HOTP) by immersing this material in various organic solvents while measuring its $T_1$ at X-band (the sample is named as $Zn_3$(HOTP)@solvent). These solvents do not affect the structure of the framework as demonstrated by X-ray diffraction and cryo-electron microscopy (Supplementary Figs. 3 – 5), yet their adsorption into the nanoscale pores (Fig. 4a) could tweak phonon dispersions of $Zn_3$(HOTP) through host-guest interactions[18]. When $Zn_3$(HOTP) is immersed in DCM, THF, and $CS_2$, its $T_1$ becomes significantly longer below 100 K and exhibits a strong temperature dependence (Fig. 4b). The spin relaxation involves a direct process and local-mode processes driven by vibrations of hydrogen bonds and Zn−O bonds, whereas the temperature-independent process is quenched (Fig. 4d, Supplementary Fig. 21, and Supplementary Table 18).

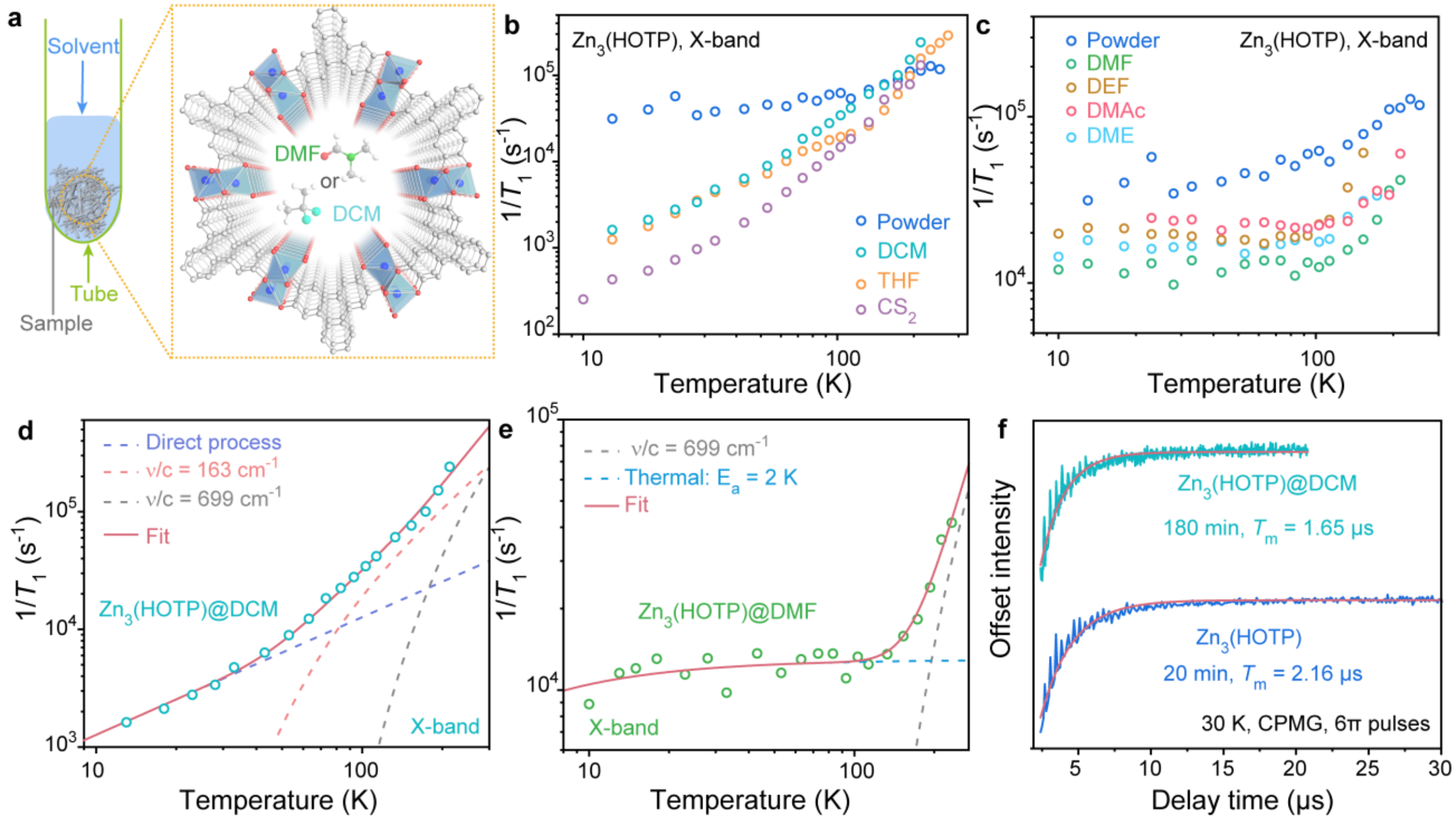

**Fig. 4. Guest modulation of spin relaxation. a,** Schematic diagram of $Zn_3$(HOTP) crystallites soaked in solvents. Solvent molecules are adsorbed into pores. **b,** Temperature dependence of $1/T_1$ for $Zn_3$(HOTP) soaked in DCM, THF, and $CS_2$ acquired at X-band. **c,** Temperature dependence of $1/T_1$ of $Zn_3$(HOTP) soaked in DMF, DEF, DMAc, and DME acquired at X-band. Error bars are omitted in **b** and **c** for clarity as they are smaller than the data point symbols (Supplementary Tables 10 – 16). **d, e,** Spin relaxation mechanisms of $Zn_3$(HOTP) soaked in (**d**) DCM and (**e**) DMF at X-band. Dash lines labeled as "Thermal" and "υ/c" represent thermally activated and local-mode processes, respectively. (**f**) CPMG echo decay traces for $Zn_3$(HOTP) powders and $Zn_3$(HOTP)@DCM collected at X-band and 30 K. Sharp spikes are electron spin echo envelope modulation signals.

Distinct from the above observation, when $Zn_3$(HOTP) is immersed in DMF, *N*,*N*-diethylformamide (DEF), *N*,*N*-dimethylacedamide (DMAc), and dimethoxyethane (DME), its $T_1$ becomes slightly longer while remaining nearly temperature-independent below 110 K, and it decreases above this temperature (Fig. 4c). Whereas the temperature-independent process persists with a slightly lower rate, the local-mode process becomes mainly driven by the stretch mode of Zn−O bond (Fig. 4e, Supplementary Fig. 22). The vibration of hydrogen bonds contributes negligibly to spin relaxation, indicating that these hydrogen-accepting and water-miscible solvents tend to disrupt hydrogen bonds between the adsorbed $H_2O$ and HOTP radicals by replacing the former in the pores. These observations demonstrate subtle tunability of phonons in $Zn_3$(HOTP) with guest molecules. Such phononic modularity is typically unfeasible for dense inorganic solids as their phonon dispersion relations are determined by crystal structures[35], yet it is easily achievable for microporous materials by harnessing host-guest interactions.

Notably, although $T_1$ of $Zn_3$(HOTP) powders at X-band is much shorter than those under higher magnetic fields or soaked in DCM, THF, and $CS_2$, $T_m$ of these samples are comparable, indicating that spin decoherence is not governed by relaxation nor relaxation-related spectral diffusion for this material (Supplementary Fig. 23). Hence, the temperature-independent process can be harnessed to improve repetition rates of quantum operations for quantum information technologies while maintaining the qubit coherence[36,37]. Typically, quantum operations need to be repeated many times, where qubits should be reset to thermal equilibrium at the beginning of each repetition[35]. Accordingly, the time required for each repetition should well exceed five times of $T_1$ to reach more than 99% relaxation[38], which often results in low experimental efficiency, unfeasibly long experimental time, and limited sensitivity for quantum sensing[35]. Thus, moderately shortened $T_1$ values are preferred to balancing the coherence of qubits with repetition rates of quantum operations. This could facilitate the integration of MQFs with other qubit platforms, e.g. superconducting qubits, that necessitate ultralow temperatures.

To demonstrate the practical significance of relaxation acceleration, we conducted nuclear-spin quantum sensing experiments for $Zn_3$(HOTP) powders ($T_1$ = 24.9 μs) and $Zn_3$(HOTP)@DCM ($T_1$ = 470 μs) at X-band and 18 K with the combination-peak electron spin echo envelope modulation (CP-ESEEM) pulse sequence. CP-ESEEM spectra of these two samples displayed signatures of $^1$H nuclear spins with comparable signal-to-noise ratio (Supplementary Fig. 24). Nonetheless, the corresponding measurements took 4.5 min and 87 min, respectively, under the same experimental parameters except for the repetition time. Similarly, applying a dynamical decoupling pulse sequence to these two samples yielded comparable improvements of $T_m$ at 30 K, yet the former sample necessitated 9 times less experimental time (Fig. 4f). Therefore, harnessing the shorter $T_1$ of $Zn_3$(HOTP) could significantly improve repetition rates for quantum sensing and dynamical decoupling while retaining their performances.

### Analysis of spin relaxation mechanism

The fast and seemingly temperature-independent spin relaxation in the powders of $Zn_3$(HOTP) at X-band is rarely observed for electron spin systems. This anomalous behavior cannot be rationalized by common molecular spin relaxation mechanisms including direct, Raman, Orbach, and local-mode processes, whose scaling factors are equal to or above 1 in our experimental temperature range (see Supplementary Information)[11,12]. Tumbling-induced spin relaxation is irrelevant as HOTP radicals are fixed by the lattice and cannot freely tumble. In addition, common semiconductor spin relaxation mechanisms, e.g. Elliot−Yafet mechanism, D'yakonov−Perl' mechanism, Bir−Aronov−Pikus mechanism, etc., are not applicable because they are derived for delocalized electrons or holes rather than small polarons[39,40].

To our knowledge, similar phenomenon has only been analyzed for nanomagnets as well as heavily doped insulators and semiconductors, where it is attributed to quantum tunneling of magnetization (QTM)[41], cross relaxation between localized spins[42,43], or mutual scattering between delocalized and localized electrons[44–46]. Nonetheless, our experimental results rule out these mechanisms for $Zn_3$(HOTP) (see Supplementary Information). First, QTM takes place between degenerate spin states[41], but HOTP radicals exhibit weak hyperfine splitting and spin-orbit coupling thereby should possess non-degenerate spin states under 0.34 T due to Zeeman splitting. Indeed, several other HOTP-based MQFs exhibit strongly temperature-dependent $T_1$[22,47], further disproving QTM as the origin of the temperature-independent spin relaxation in $Zn_3$(HOTP). Second, the cross relaxation rate should scale with the electron spin density. We prepared a sample of $Zn_3$(HOTP) where 0.012% of HOTP ligands form small polarons ($Zn_3$(HOTP)-L, Supplementary Table 5). This ratio is approximately two orders of magnitude lower, but the sample still displayed nearly temperature-independent $T_1$ between 30 K and 80 K (Supplementary Fig. 25 and Supplementary Table 7), indicating that the anomalous spin relaxation does not stem from cross relaxation. Third, the mutual scattering mechanism should relate to mobility of delocalized electrons as well as densities of both delocalized and localized electrons in $Zn_3$(HOTP), which should not be sensitive to redox-innocent and non-coordinating solvents like DCM. Indeed, soaking $Zn_3$(HOTP) in DCM barely changes the density of small polarons and large polarons (Supplementary Fig. 26; Supplementary Table 5) and exposing its single crystals to DCM vapor does not alter room-temperature electrical conductivity (Supplementary Figs. 15 and 16, Supplementary Table 4). These experimental results together confirm that DCM does not tweak the charge mobility. Accordingly, the adsorption of this solvent should not modulate spin-spin interaction (cross relaxation) or mutual scattering between delocalized and localized electrons, excluding these two mechanisms.

As the adsorption of DCM could subtly tweak phonons of $Zn_3$(HOTP) through host-guest interactions, the seemingly temperature-independent spin relaxation likely stems from spin-phonon coupling. Indeed, it is consistent with the thermally activated process where $T_1$ could be temperature-independent when the activation energy barrier ($E_a$, in the unit of temperature) is much lower than the lowest experimental temperature. Hence, we revise Equation 1 as follows:

$$\frac{1}{T_1} = A_{Therm}\frac{2\tau_c}{1+\omega^2{\tau_c}^2} + A_{Dir}T + \sum_i A_{Loc,i}\frac{e^{h\nu_i/k_BT}}{\left(e^{h\nu_i/k_BT}-1\right)^2} \quad \text{......................(2)}$$

Here, $A_{Therm}$, $\tau_c$, and $\omega$ represent the pre-factor, correlation time, and electron spin Larmor frequency of the thermally activated process, respectively, with $\tau_c = \tau_c^0 \exp(E_a/k_BT)$ where $\tau_c^0$ is a preexponential factor[11]. Fitting the temperature dependence of $T_1$ with Equation 2 revealed $E_a$ of 5 K for $Zn_3$(HOTP) powder at 0.34 T, which drops to 2 K upon soaking this material in DMF, DEF, and DME (Fig. 3d and Fig. 4e; Supplementary Fig. 27 and Supplementary Table 19). Although a low-barrier thermally activated process could account for the experimental data, its underlying microscopic mechanism remains unclear. Key questions persist: what is the nature of the barrier-crossing event? Why is the spin relaxation rate sensitive to both magnetic fields and pore-filling solvents? A new mechanism is demanded to explain the anomalous spin relaxation behavior of $Zn_3$(HOTP).

## Spin-phonon polarons

Motivated by the intrinsic structural chirality as well as the unique field- and guest-tunable spin relaxation behaviors of $Zn_3$(HOTP), we introduce an additional relaxation channel assisted by spin-phonon polarons, which is described by the following non-adiabatic microscopic model:

$$\hat{H} = \frac{\Delta}{2}\hat{\sigma}_z + \hat{H}_\mathrm{B} + \left(\alpha_1\hat{Q}_\mathrm{sl} + \alpha_2\hat{Q}_\mathrm{sl}^2\right)\hat{\sigma}_x + \gamma_1\hat{Q}_\mathrm{sp}\hat{\sigma}_z \quad \text{....................(3)}$$

Here, $\hat{\sigma}_z$ and $\hat{\sigma}_x$ are Pauli matrices; $\alpha_1$, $\alpha_2$, and $\gamma_1$ are spin-phonon coupling constants. $\gamma_1$ is proportional to the external magnetic field strength ($B$) originated from the vibrational dependence of the Landé $g$-factor[48,49]. The four terms describe the electronic Zeeman splitting, lattice phonon bath, linear and quadratic transverse spin-phonon coupling, as well as longitudinal spin-phonon (polaron-like) interaction, respectively. The spin system with Zeeman splitting $\Delta$ couples with two bands of phonons: the collective mode $\hat{Q}_\mathrm{sl}$ drives conventional spin-lattice relaxation, e.g. direct, Raman, and local-mode processes (see Supplementary Information), while an additional mode $\hat{Q}_\mathrm{sp}$ asymmetrically couples with different spin states, enabling the formation of a spin-phonon polaron (Fig. 5a). Notice that the polaron displacement is proportional to $\gamma_1$, which is again proportional to $B$, and that the sign of $\gamma_1$ reflects the handedness of the chirality[9,50].

A straightforward Fermi's Golden Rule calculation (see Supplementary Information) indicates that the spin-phonon polaron modulates the spin relaxation rate with a factor of $\exp\left[-\frac{(\Delta-E_r)^2}{4k_BTE_r}\right]$, analogous to the Marcus theory[51]. The reorganization energy $E_r$ depends on both the phonon spectral density and $B$. In particular, $E_r$ shows a strong field dependence as $E_r \propto {\gamma_1}^2 \propto B^2$. In the high-temperature limit, the quantum Fermi's Golden Rule reduces to a classical thermally activated transition with $E_a = (\Delta - E_r)^2/4E_r$, indicating that the thermally activated spin relaxation process is a natural consequence of the spin-phonon polaron-assisted relaxation. Notably, the spin-phonon polaron modulation is independent of the structural chirality as $E_r$ is quadratic to $\gamma_1$. As a result, this mechanism is applicable to the racemate of $Zn_3$(HOTP) used in the experiments.

This microscopic model explains the abovementioned fast and nearly temperature-independent spin relaxation. As the magnetic field increases, both $\Delta$ and spin-phonon polaron displacement scale linearly with $B$, whereas $E_r$ is quadratic to $B$ (Fig. 5b). At lower magnetic fields, the spin-state transitions are more accessible because little thermal fluctuation is required to reach the potential crossing. As a result, the spin-phonon polaron-assisted relaxation channel is open and dominant, leading to efficient spin relaxation with weak temperature dependence (see Supplementary Information). This is likely the situation for $Zn_3$(HOTP) powders and those soaked in DMF, DEF, DMAc, and DME at X-band. The temperature dependencies of $T_1$ observed for these samples can be well fitted by the rate function derived from Equation 3 (Fig. 5c and d; Supplementary Fig. 28).

This model also explains the field- and guest-induced suppression of the above spin relaxation behavior. Raising the magnetic field from 0.34 T to 1.22 T (or 3.34 T) raises $E_r$. Changing guest molecules in the pores might alter the vibrational environment around the radical, which varies the spin-phonon coupling strength and density of states thereby enhancing $E_r$. In either case, $E_r$ could become too large to be overcome by thermal fluctuations, effectively suppressing the spin-phonon polaron-assisted relaxation channel. As a result, the conventional direct and local-mode processes dominate the relaxation dynamics. This is likely the situation for $Zn_3$(HOTP) powders at Q-band and W-band as well as those soaked in DCM, THF, and $CS_2$ at X-band. The temperature dependencies of $T_1$ observed for these samples can be successfully fitted using the standard rate expressions for conventional spin relaxation mechanisms (Fig. 3e and f; Fig. 4d; Supplementary Fig. 21). As the spin-phonon polaron-assisted relaxation channel is sensitive to magnetic field and phonon environment, the observation of this quasiparticle in other chiral materials might require spin dynamic characterization at relatively weak magnetic fields and well-controlled atmosphere.

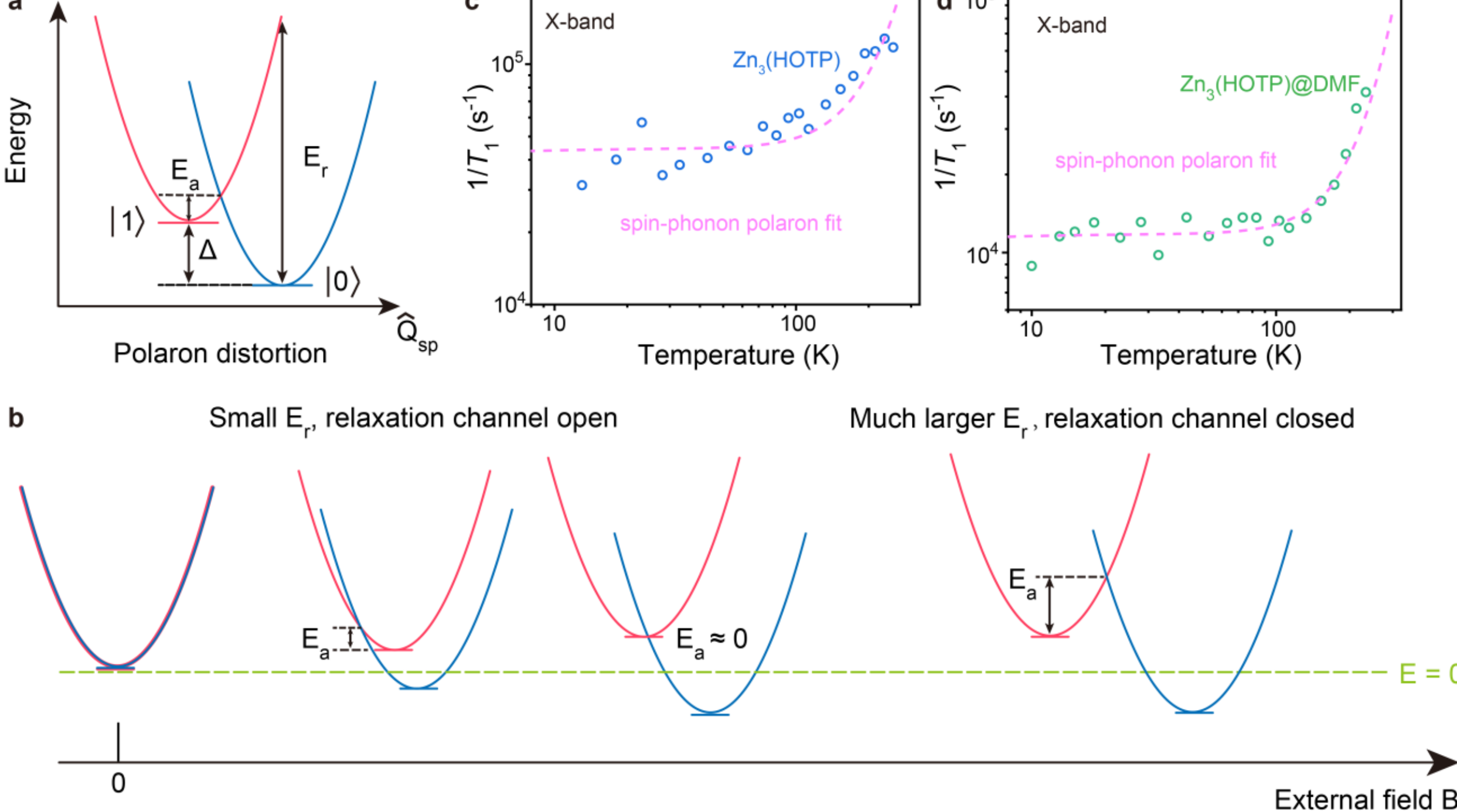

**Fig. 5**. **Spin-phonon polaron-assisted spin relaxation. a**, Potential energy surfaces of different spin states ($|0\rangle$ and $|1\rangle$) of a $S$ = 1/2 system as a function of polaron distortion coordinate $\hat{Q}_{\mathrm{sp}}$. **b**, Mechanism for the magnetic-field-controlled activation and suppression of the spin-phonon polaron-assisted relaxation channel. **c,d,** Fitting of the temperature dependence of spin relaxation rates for (**c**) $Zn_3$(HOTP) powders and (**d**) $Zn_3$(HOTP)@DMF acquired at X-band based on rate functions derived from Equation 3 (see Supplementary Information).

The foregoing results provide the first experimental evidence of spin-phonon polarons in a chiral and microporous MQF, $Zn_3$(HOTP). They likely emerge from chirality-induced asymmetric spin-phonon coupling and cause anomalous and tunable spin relaxation behaviors by introducing an additional relaxation channel. As a result, the seemingly unexpected temperature-independent feature of spin relaxation rate can be then explained by the existence of barrierless spin-phonon polaron state. The formation of these quasiparticles should not be limited to MQFs—they should be able to form in chiral polymers, soft inorganic solids (e.g. chiral halide perovskites), and artificial structures under appropriate (likely weak) magnetic fields. Further investigations of their dynamics and their influence on electron (or spin) transport could reveal insights into the interplay between structural chirality and electron spin degree of freedom and may shed light on the mechanism of CISS effect. In addition, spin-phonon polarons could facilitate relaxation while maintaining coherence for electron spin qubits, which might occur in solid-state paramagnetic defects and semiconductor quantum dot qubits as well. Harnessing this unique property could shorten the time needed for qubit initialization thereby improving repetition rates of quantum operations, leading to new strategies for high-efficiency spin-based quantum computing and quantum sensing.

# Methods

## Synthesis of $Zn_3$(HOTP)

In a $N_2$-filled glovebox, HHTP (0.0370g, 0.1141 mmol; home-made compound) was weighed into a 23 mL Teflon liner and then transferred out of the glove box. Zn(acetate)$_2$·2$H_2O$ (0.2003 g, 0.9148 mmol), 5 mL of deionized water, and 0.5 mL of *N*-methyl-2-pyrrolidone were added into the Teflon liner successively. The mixture was sonicated for 30 min and then transferred into an autoclave. It was heated in an oven at 120 °C for 24 hours and then cooled naturally to room temperature, yielding blue powders. The product was collected by filtration and washed with water, ethanol, and acetone. Elemental analysis calculated for $Zn_3$(HOTP)($H_2O$)$_3$·4$H_2O$ ($Zn_3C_{18}H_{20}O_{13}$): C, 33.76%; H, 3.15%; N, 0.0%; Found: C, 33.79%; H, 3.47%; N, <0.3%.

## cRED data collection, procession, and structure solution

Crystals of as-synthesized $Zn_3$(HOTP) were dispersed in ethanol and ultrasonication for 5 min. A droplet of suspension was then transferred on a copper grid with carbon film. The cRED data were collected on 200kV JEOL JEM-2100 transmission electron microscope equipped with a quad hybrid pixel detector (Timepix, 512 × 512 pixels, pixel size 55 μm, Amsterdam Sci. Ins.). Before data collection, the sample was cooled down to 96 K by using Gatan cryo-transfer tomography holder. During the data collection, the goniometer was rotated continuously while the selected area electron diffraction (ED) patterns were captured from the crystal simultaneous by using the software of *instamatic*[1]. In order to balance the intensity of electron diffraction and resolution, all the ED patterns were recorded under the spot size 3 with the exposure time 0.5 s. The 3D reciprocal lattice was reconstructed by the software REDp[2], which was very useful for indexing and obtaining the reflection conditions.

For the structure solution of $Zn_3$(HOTP), the crystals could be indexed with hexagonal symmetry, with lattice parameters of $a$ = 21.166(3) Å, $c$ = 3.3046(7) Å in space group $P6_3$ (Supplementary Fig. 6, reflection condition: *000l, l=2n*). It should be noted that along the crystallographic $c$ axis displays satellite reflections with q* ~ 2/7c*. The X-ray crystallography software package *XDS*[3] was used for data processing to estimate integrated diffraction intensities. All non-hydrogen atomic positions in $Zn_3$(HOTP) framework can be located directly by ab initial method. The SHELXL was used for structure refinement using electron scattering factors for all the atoms.

## CW-EPR spectroscopy

X-band (9.6 GHz) CW-EPR spectroscopy was performed on Bruker E500 spectrometer at the Instrumentation and Service Center for Molecular Sciences, Westlake University. A standard BDPA radical sample was used to calibrate the magnetic field, which revealed +0.076 mT correction[4]. Experimental parameters including modulation amplitude, microwave power, and temperature were set to 0.01 mT, 0.1 mW, and 296 K, respectively.

Q-band (34 GHz) CW-EPR spectroscopy was performed on a Bruker E580 spectrometer equipped with an Oxford ESR910 liquid helium cryostat at East China Normal University. A standard BDPA radical sample was used to calibrate magnetic field, which revealed +3.05 mT correction.

Millimeter-wave (214 GHz) CW-EPR spectroscopy was conducted at High Magnetic Field laboratory, Hefei Institutes of Physical Science, Chinese Academy of Sciences. The magnetic field was calibrated by a standard DPPH sample, which revealed −9.01 mT correction.

CW-EPR spectra were fitted by EasySpin 6.0.05 in MATLAB R2023b[5].

**Pulse EPR spectroscopy**

X-band (9.6 GHz) pulse EPR spectroscopy was performed on a CIQTEK EPR100 spectrometer equipped with EPR-VTS-L-D1-PWC closed-loop helium cryostat and Lakeshore336 temperature controller at the Instrumentation and Service Center for Molecular Sciences, Westlake University. A standard DPPH radical sample was used to calibrate magnetic field, which revealed −0.351 mT correction.

Q-band (34 GHz) pulse EPR spectroscopy was performed on a Bruker E580 spectrometer equipped with an Oxford ESR910 liquid helium cryostat at East China Normal University. A standard BDPA radical sample was used to calibrate magnetic field, which revealed +3.05 mT correction.

W-band (94 GHz) pulse EPR spectroscopy was conducted on a CIQTEK EPR-W900 spectrometer at Chinainstru & Quantumtech (Hefei) Co.,Ltd.

Echo-detected field sweep (EDFS) experiments

The X-band EDFS spectra was collected using a two pulse Hahn echo sequence (π/2 –τ – π– τ – echo) with 1000 μs repetition time, 100 shots per point, and 1024 data points. π/2 and π pulses were applied with lengths of 128 ns and 256 ns, respectively. Two-step phase cycling was employed with pulse phases of (+x, +x) and (−x, +x) to cancel background drift and the defense pulse.

Nutation experiments

Nutation experiments were performed with a three-pulse sequence (nutation pulse – T – π/2 – τ – π – τ – echo) with 512 data points in various microwave attenuations (0, 3, 6, 9, 12, 15, and 18 dB) at X-band and 295 K. The duration of the nutation pulse began at 6 ns and increased by 2 ns with each subsequent step. T and τ were set to 400 and 200 ns, respectively. The background noise was canceled by four-step phase cycling with pulse phases of (+x, −x, +x) (+x, +x, +x) (−x, −x, +x) and (−x, +x, +x). Fast Fourier transform (FFT) in Origin 2023b was used to transform nutation curves from the time domain to the frequency domain after baseline correction with cubic polynomials, apodization with the Hamming window function, and zero-filling. Rabi frequencies recorded at the frequency of the peak after FFT are different at different microwave attenuations, which are plotted against the ratio between the magnetic field of the output microwave (Bout, after attenuation) and the input microwave ($B_{in}$, before attenuation), which is $10^{\frac{-Attenuation}{20\,dB}}$ with Attenuation the microwave attenuation in the unit of dB.

Hahn echo decay measurements

Phase memory time ($T_m$) was characterized by the Hahn echo decay pulse sequence (π/2 – τ – π – τ – echo). The length of π pulse was 32 ns for X-band and Q-band EPR, and it varied for W-band EPR at each temperature. The background noise was canceled by two-step phase cycling with pulse phases of (+x, +x) and (−x, +x). The Hahn echo was monitored under various τ (Supplementary Figs. 29 – 39). The resulting Hahn echo decay curve was fitted by the following exponential decay function to extract $T_m$ (Supplementary Tables 6 – 16).

$$I = A_{1,m}e^{-\frac{2\tau}{T_m}} + A_{2,m}e^{-\frac{2\tau}{T_m^S}} + I_{0,m}$$

where $I$ is echo intensity, $A_{1,m}$ and $A_{2,m}$ are pre-factors, $T_m^S$ describes a fast relaxation process arising from spectral diffusion, and $I_{0,m}$ accounts for the baseline drift.

Inversion recovery measurements

Spin relaxation time ($T_1$) was collected by an inversion recovery sequence (π – T– π/2 – τ – π – τ – echo). The length of π pulse was 32 ns for X-band and Q-band EPR, and it varied for W-band EPR at each temperature. The background noise was canceled by four-step phase cycling with pulse phases of (+x, −x, +x) (+x, +x, +x) (−x, −x, +x) and (−x, +x, +x). For variable-temperature pulse EPR spectra, samples were cooled below 10 K for at least 2 hours before conducting spin dynamics experiments. Temperature was controlled by balancing liquid helium flow rate and the power of a heater. The temperature of sample was monitored by comparing $T_1$ values obtained from consecutive inversion recovery experiments — the thermal equilibrium is assumed to be reached when the $T_1$ value changes less than 1%. It typically takes 20 to 30 minutes to reach the thermal equilibrium.

The echo was monitored under various T (Supplementary Figs. 29 – 39). Inversion recovery curves were fitted by the following biexponential decay function to extract $T_1$ (Supplementary Tables6 – 16).

$$I = A_1 e^{\frac{-T}{T_1}} + A_2 e^{\frac{-T}{T_s}} + I_0$$

where $I$ is echo intensity, $A_1$ and $A_2$ are pre-factors, $T_s$ describes a fast relaxation process arising from spectral diffusion, and $I_0$ accounts for the baseline drift.

CPMG measurements

The phase memory time was prolonged by CPMG sequence, which can be represented as ($\pi_x/2$ – [τ – $\pi_y$ – 2τ – $\pi_y$ – τ]$_n$ – echo), at X-band and 30 K. The length of π pulse was 32 ns. In CPMG experiments, phase cycling is needed to guarantee reliable coherence-time measurements. Here, each $\pi_y$ pulse has two phase choices: +Y and –Y. The results from all possible phase configurations of $\pi_y$ pulses were summed, except that the first $\pi_y$ pulse's phase was fixed as +Y. The first $\pi_x/2$ pulse was phase cycled like in Hahn echo decay measurements. The CPMG echo was monitored under various τ[6–8]. $T_m$ was extracted by fitting the relationship between echo intensity and 4nτ with a mono-exponential decay function.

Combination–peak electron spin echo envelope modulation (CP-ESEEM) measurements

CP-ESEEM spectroscopy was performed by X-band with a four-pulse sequence (π/2 – τ – π/2 – T – π – T – π/2 – τ – echo), where the π pulse length was 32 ns and the data points was 512. T started at 400 ns and was incremented 8 ns per step; τ was fixed at 120 ns. The background noise, unwanted echoes, and the defense pulse were canceled by eight-step phase cycling, where the pulse phases are (+x, +x, +x, +x) (−x, +x, +x, +x) (+x, −x, +x, +x) (−x, −x, +x, +x) (+x, +x, +x, −x) (−x, +x, +x, −x) (+x, −x, +x, −x) (−x, −x, +x, −x). The resulting curve was an oscillatory time-domain CP-ESEEM spectrum, which was transformed into a frequency-domain spectrum with background correction using polynomial fitting, apodization with the Hamming window function, zero-padding, and fast Fourier transformation.

## Data availability

The data supporting our findings are included in the article and supplementary files. Additional data generated during the study are available from the corresponding author upon request.

## References

1. Göhler, B. *et al.* Spin selectivity in electron transmission through self-assembled monolayers of double-stranded DNA. *Science* **331**, 894–897 (2011).

2. Bhowmick, D. K. *et al.* Spin-induced asymmetry reaction—the formation of asymmetric carbon by electropolymerization. *Sci. Adv.* **8**, eabq2727 (2022).

3. Lu, H. *et al.* Spin-dependent charge transport through 2D chiral hybrid lead-iodide perovskites. *Sci. Adv.* **5**, eaay0571 (2019).

4. Naaman, R., Paltiel, Y. & Waldeck, D. H. Chiral molecules and the electron spin. *Nat. Rev. Chem.* **3**, 250–260 (2019).

5. Yang, S.-H., Naaman, R., Paltiel, Y. & Parkin, S. S. P. Chiral spintronics. *Nat. Rev. Phys.* **3**, 328–343 (2021).

6. Chiesa, A. *et al.* Chirality-induced spin selectivity: an enabling technology for quantum applications. *Adv. Mater.* **35**, e2300472 (2023).

7. Bloom, B. P. *et al.* Asymmetric reactions induced by electron spin polarization. *Phys. Chem. Chem. Phys.* **22**, 21570–21582 (2020).

8. Bloom, B. P., Paltiel, Y., Naaman, R. & Waldeck, D. H. Chiral induced spin selectivity. *Chem. Rev.* **124**, 1950–1991 (2024).

9. Fransson, J. Vibrational origin of exchange splitting and chiral-induced spin selectivity. *Phys. Rev. B* **102**, 235416 (2020).

10. Wang, Z.-W. *et al.* Polaron spin states in chiral systems. *Phys. Rev. B* **110**, 085430 (2024).

11. S. S. Eaton, G. R. Eaton, *EPR Spectroscopy: Fundamentals and Methods Ch.9*. (John Wiley & Sons Ltd, Hoboken, 2018)

12. Lunghi, A. *Computational Modelling of Molecular Nanomagnets Ch.6*. (Springer Cham, Switzerland, 2023).

13. Xu, J. *et al.* How spin relaxes and dephases in bulk halide perovskites. *Nat. Commun.* **15**, 188 (2024).

14. Jee, B., Hartmann, M. & Pöppl, A. $H_2$, $D_2$ and HD adsorption upon the metal-organic framework $[Cu_{2.97}Zn_{0.03}(btc)_2]_n$ studied by pulsed ENDOR and HYSCORE spectroscopy. *Mol. Phys.* **111**, 2950–2966 (2013).

15. Sun, L. *et al.* Room-temperature quantitative quantum sensing of lithium ions with a radical-embedded metal–organic framework. *J. Am. Chem. Soc.* **144**, 19008–19016 (2022).

16. Oanta, A. K. *et al.* Electronic spin qubit candidates arrayed within layered two-dimensional polymers. *J. Am. Chem. Soc.* **145**, 689–696 (2023).
17. Yu, C.-J. *et al.* Spin and phonon design in modular arrays of molecular qubits. *Chem. Mater.* **32**, 10200–10206 (2020).
18. Vujević, L. *et al.* Improving the molecular spin qubit performance in zirconium MOF composites by mechanochemical dilution and fullerene encapsulation. *Chem. Sci.* **14**, 9389–9399 (2023).
19. Gong, W., Chen, Z., Dong, J., Liu, Y. & Cui, Y. Chiral metal–organic frameworks. *Chem Rev* **122**, 9078–9144 (2022).
20. Orihashi, K. *et al.* Spin-polarized radicals with extremely long spin–lattice relaxation time at room temperature in a metal–organic framework. *J. Am. Chem. Soc.* **145**, 27650–27656 (2023).
21. Zhou, A., Sun, Z. & Sun, L. Stable organic radical qubits and their applications in quantum information science. *The Innovation*, **5**, 100662 (2024).
22. Zhou, A. *et al.* Phononic modulation of spin-lattice relaxation in molecular qubit frameworks. *Nat. Commun.* **15**, 10763 (2024).
23. Holstein, T. Studies of polaron motion Part II. The "small" polaron. *Ann. Phys.* **8**, 343–389 (1959).
24. Leggett, A. J. *et al.* Dynamics of the dissipative two-state system. *Rev. Mod. Phys.* **59**, 1–85 (1987).
25. Würger, A. Strong-coupling theory for the spin-phonon model. *Phys. Rev. B* **57**, 347–361 (1998).
26. Garwoła, J. & Segal, D. Open quantum systems with noncommuting coupling operators: An analytic approach. *Phys. Rev. B* **110**, 174304 (2024).
27. Choi, J. Y. *et al.* 2D conjugated metal-organic framework as a proton-electron dual conductor. *Chem* **9**, 143–153 (2023).
28. Huang, W., Xiao, Y., Xia, F., Chen, X. & Zhai, T. Second harmonic generation control in 2d layered materials: status and outlook. *Adv. Funct. Mater.* **34**, 2310726 (2024).
29. Jähnigen, S. Vibrational circular dichroism spectroscopy of chiral molecular crystals: insights from theory. *Angew. Chem. Int. Ed.* **62**, e202303595 (2023).
30. Zhang, K. J. *et al.* High-resolution structure of $Zn_3(HOTP)_2$ (HOTP = hexaoxidotri-phenylene), a three-dimensional conductive MOF. *Chem. Sci.* **16**, 12416–12420 (2025).

31. Skorupskii, G. *et al.* Porous lanthanide metal–organic frameworks with metallic conductivity. *Proc. Natl. Acad. Sci.* **119**, e2205127119 (2022).

32. Sitaram, V., Sharma, A., Bhat, S. V., Mizoguchi, K. & Menon, R. Electron spin resonance studies in the doped polyaniline PANI-AMPSA: Evidence for local ordering from linewidth features. *Phys Rev B* **72**, 035209 (2005).

33. Chen, Q., Adeniran, O., Liu, Z.-F., Zhang, Z. & Awaga, K. Graphite-like charge storage mechanism in a 2D π–d conjugated metal–organic framework revealed by stepwise magnetic monitoring. *J. Am. Chem. Soc.* **145**, 1062–1071 (2023).

34. Franchini, C., Reticcioli, M., Setvin, M. & Diebold, U. Polarons in materials. *Nat. Rev. Mater.* **6**, 560–586 (2021).

35. Bienfait, A. *et al.* Controlling spin relaxation with a cavity. *Nature* **531**, 74–77 (2016).

36. Tornow, C., Kanazawa, N., Shanks, W. E. & Egger, D. J. Minimum quantum run-time characterization and calibration via restless measurements with dynamic repetition rates. *Phys. Rev. Appl.* **17**, 064061 (2022).

37. Jones, J. A. Controlling NMR spin systems for quantum computation. *Prog. Nucl. Magn. Reson. Spectrosc.* **140**, 49–85 (2024).

38. Jackson, C. E. *et al.* Impact of counter ion methyl groups on spin relaxation in $[V(C_6H_4O_2)_3]^{2-}$. *J. Phys. Chem. C* **126**, 7169–7176 (2022).

39. Wu, M. W., Jiang, J. H. & Weng, M. Q. Spin dynamics in semiconductors. *Phys. Rep.* **493**, 61–236 (2010).

40. Schott, S. *et al.* Polaron spin dynamics in high-mobility polymeric semiconductors. *Nat. Phys.* **15**, 814–822 (2019).

41. Gatteschi, D. & Sessoli, R. Quantum tunneling of magnetization and related phenomena in molecular materials. *Angew. Chem. Int. Ed.* **42**, 268–297 (2003).

42. Jarmola, A., Acosta, V. M., Jensen, K., Chemerisov, S. & Budker, D. Temperature- and magnetic-field-dependent longitudinal spin relaxation in nitrogen-vacancy ensembles in diamond. *Phys. Rev. Lett.* **108**, 197601 (2012).

43. Lu, Y. *et al.* Rational construction of layered two-dimensional conjugated metal–organic frameworks with room-temperature quantum coherence. *J. Am. Chem. Soc.* **147**, 8778–8784 (2025).

44. Völkel, G. & Brunner, W. Electron spin-lattice relaxation of trapped EPE active centres in $O_2^-$ contaminated polyacetylene. *Phys. status solidi (a)* **94**, 673–678 (1986).

45. Song, Y., Chalaev, O. & Dery, H. Donor-driven spin relaxation in multivalley semiconductors. *Phys Rev Lett* **113**, 167201 (2014).

46. Fujita, Y. *et al.* Temperature-independent spin relaxation in heavily doped n-type germanium. *Phys Rev B* **94**, 245302 (2016).

47. Sun, Z. *et al.* Ultralong room-temperature qubit lifetimes of covalent organic frameworks. *J. Am. Chem. Soc.* **147**, 31930–31939 (2025).

48. Vleck, J. H. V. Paramagnetic relaxation times for titanium and chrome alum. *Phys. Rev.* **57**, 426–447 (1940).

49. Lunghi, A. & Sanvito, S. How do phonons relax molecular spins? *Sci. Adv.* **5**, eaax7163 (2019).

50. Teh, H.-H., Dou, W. & Subotnik, J. E. Spin polarization through a molecular junction based on nuclear Berry curvature effects. *Phys. Rev. B* **106**, 184302 (2022).

51. Nitzan, A. *Chemical Dynamics in Condensed Phases: Relaxation, Transfer and Reactions in Condensed Molecular Systems* (Oxford University Press, Oxford, 2006).

## Methods references

1. Lu, H. *et al.* Spin-dependent charge transport through 2D chiral hybrid lead-iodide perovskites. *Sci. Adv.* **5**, eaay0571 (2019).

2. Naaman, R., Paltiel, Y. & Waldeck, D. H. Chiral molecules and the electron spin. *Nat. Rev. Chem.* **3**, 250–260 (2019).

3. Yang, S.-H., Naaman, R., Paltiel, Y. & Parkin, S. S. P. Chiral spintronics. *Nat. Rev. Phys.* **3**, 328–343 (2021).

4. Daniel, D. T. *et al.* Multimodal investigation of electronic transport in PTMA and its impact on organic radical battery performance. *Sci. Rep.* **13**, 10934 (2023).

5. Stoll, S. & Schweiger, A. EasySpin, a comprehensive software package for spectral simulation and analysis in EPR. *J. Magn. Reson.* **178**, 42–55 (2006).

6. Gemperle, C., Aebli, G., Schweiger, A. & Ernst, R. R. Phase cycling in pulse EPR. *J. Magn. Reson. (1969)* **88**, 241–256 (1990).

7. Du, J. *et al.* Preserving electron spin coherence in solids by optimal dynamical decoupling. *Nature* **461**, 1265–1268 (2009).

8. Lombardi, F. *et al.* Dynamical nuclear decoupling of electron spins in molecular graphenoid radicals and biradicals. *Phys. Rev. B* **101**, 094406 (2020).

## Acknowledgements

This work was supported by the National Natural Science Foundation of China (No. 22273078, No. 22361142829, No. 22371121), Zhejiang Provincial Natural Science Foundation (No. XHD24B0301, No. XHD23B0301), the Hangzhou Municipal Funding Team of Innovation (TD2022004), and Scientific Research Project of Westlake University (No. WU2024B025). A portion of this work was carried out at the Synergetic Extreme Condition User Facility (SECUF, https://cstr.cn/31123.02.SECUF), the Electron Spin Resonance System at the Steady High Magnetic Field Facility, CAS (https://cstr.cn/31125.02.SHMFF), and the 17B beamline of Shanghai Synchrotron Radiation Facility (SSRF). T.R. is supported by the College of Liberal Art and Sciences at Arizona State University. Work in the Dincă group was funded by the Department of Energy, Basic Energy Sciences (DE-SC0023288). Computational resources were provided by the Westlake HPC Center. A.Z. and L.S. acknowledge Dr. Fushan Geng and Prof. Bingwen Hu for assistance with Q-band EPR spectroscopy, Zengwen Wang and Prof. Zhenxing Wang for assistance with preliminary millimeter-wave EPR experiments, and Prof. Shuai Yuan for assistance with preliminary single-crystal X-ray crystallography. A.Z. thanks Dr. Qike Jiang, Danyu Gu, Dr. Zhong Chen, Zhongwei Yang, Yunlong Fan, and Tongyang Zhao for assistance with material characterization and thanks Dr. Ling Zhang, Hao Chen, and Tongyang Zhao for assistance with material synthesis. A.Z. and L.S. thank the Instrumentation and Service Center for Molecular Sciences and the Instrumentation and Service Center for Physical Sciences at Westlake University for facility support and technical assistance. A. Z. and L.S. thank Dr. Zhongyue Zhang, Prof. Congjun Wu, Prof. Shen Zhou, Jiayi Huang, and Prof. Hongfei Wang for helpful discussions. R.B. and W.D. acknowledge the helpful discussions with Prof. Yijing Yan.

## Author contributions

L.S., W.D., A.Z., and R.B. conceived the idea, designed experiments, and oversaw the project. A.Z. conducted material synthesis, EPR characterization, and vibrational spectroscopy. R.B. performed theoretical analysis. Z.Z., X.T., and J.L. conducted structural determination. L.Y, M.D., performed preliminary synthesis and material characterization. D.L. and S.L. conducted phonon DOS calculation. Y.W. conducted single-crystal device fabrication and electrical characterization. M.T. and F.J. assisted with Raman spectroscopy. W.N. assisted with dynamical decoupling and quantum sensing experiments. H.S. assisted with analysis of magnetic properties. X.M. assisted with *in situ* 2D XRD. J.G. and J.D. assisted with TEM imaging and structural analysis. X.Z. and Z.S. assisted with W-band EPR spectroscopy. Z.Z. and W.T. performed millimeter-wave EPR spectroscopy. J.F. assisted with material synthesis and EPR analysis. T.R. performed preliminary X-band EPR characterization. L.S., W.D., A.Z., and R.B. co-wrote the manuscript. All authors have given approval to the final version of the manuscript.

## Competing Interests

The authors declare that they have no competing interests.

## Materials & correspondence

Correspondence and requests for materials should be addressed to Lei Sun and Wenjie Dou.